\documentclass[12pt,twoside]{article}
\usepackage[mathscr]{eucal}
\usepackage{amsmath,amsfonts,amssymb,amsthm}
\usepackage{cite}
\usepackage[matrix,arrow]{xy}
\usepackage{amsthm,amsmath,latexsym,amssymb,amsfonts,amsbsy}
\usepackage{wrapfig}
\usepackage{graphicx}
\usepackage{float}

\usepackage[breaklinks]{hyperref}
\usepackage{amsmath}
\usepackage{amssymb}
\usepackage{braket}
\usepackage[T1]{fontenc}

\usepackage{etoolbox}
\usepackage{multido}

\makeatletter
\def\@@bfil{\leaders \vrule \@height \ht\z@ \@depth \z@ \hfill}
\def\@bLfil{\@@bfil}
\def\@bRfil{\@@bfil}
\def\resetbraceratio{\gdef\@bLfil{\@@bfil}\gdef\@bRfil{\@@bfil}}
\def\setbraceratio#1#2{
  \let\@bLfil\relax
  \multido{\iA=1+1}{#1}{\gappto\@bLfil{\@@bfil}}
  \let\@bRfil\relax
  \multido{\iA=1+1}{#2}{\gappto\@bRfil{\@@bfil}}
}
\def\upbracefill{$\m@th\setbox\z@\hbox{$\braceld$}\bracelu\@bLfil\bracerd\braceld\@bRfil\braceru$}
\def\downbracefill{$\m@th\setbox\z@\hbox{$\braceld$}\braceld\@bLfil\braceru\bracelu\@bRfil\bracerd$}
\makeatother

\usepackage{tikz,pgf}
\usetikzlibrary{shapes}
\usetikzlibrary{calc}
\usetikzlibrary{decorations.pathmorphing}
\usetikzlibrary{decorations.pathreplacing,shapes.misc}
\usetikzlibrary{positioning}
\usetikzlibrary{arrows}
\usetikzlibrary{decorations.markings}

\def\be{\begin{equation}}
\def\ee{\end{equation}}
\def\ba{\begin{array}}
\def\ea{\end{array}}
\def\ll{\left\lgroup}
\def\rr{\right\rgroup}

\def\1{\tilde{1}}
\def\2{\tilde{2}}
\def\3{\tilde{3}}

\newdimen\tableauside\tableauside=.5ex
\newdimen\tableaurule\tableaurule=0.4pt
\newdimen\tableaustep
\def\phantomhrule#1{\hbox{\vbox to0pt{\hrule height\tableaurule
width#1\vss}}}
\def\phantomvrule#1{\vbox{\hbox to0pt{\vrule width\tableaurule
height#1\hss}}}
\def\sqr{\vbox{%
  \phantomhrule\tableaustep

\hbox{\phantomvrule\tableaustep\kern\tableaustep\phantomvrule\tableaustep}%
  \hbox{\vbox{\phantomhrule\tableauside}\kern-\tableaurule}}}
\def\squares#1{\hbox{\count0=#1\noindent\loop\sqr
  \advance\count0 by-1 \ifnum\count0>0\repeat}}
\def\tableau#1{\vcenter{\offinterlineskip
  \tableaustep=\tableauside\advance\tableaustep by-\tableaurule
  \kern\normallineskip\hbox
    {\kern\normallineskip\vbox
      {\gettableau#1 0 }%
     \kern\normallineskip\kern\tableaurule}%
  \kern\normallineskip\kern\tableaurule}}
\def\gettableau#1 {\ifnum#1=0\let\next=\null\else
  \squares{#1}\let\next=\gettableau\fi\next}

\newcommand{\p}{\partial}

\newcommand{\bref}[1]{\textbf{\ref{#1}}}

\def\be{\begin{eqnarray}}
\def\ee{\end{eqnarray}}
\def\bew{\begin{eqnarray*}}
\def\eew{\end{eqnarray*}}

\def\p{\partial}

\def\l[{\phantom.[}

\numberwithin{equation}{section} \makeatletter
\@addtoreset{equation}{section}

\def\be{\begin{equation}}
\def\ee{\end{equation}}
\def\ba{\begin{array}}
\def\ea{\end{array}}

\begin{document}

\begin{flushright}
FIAN-TD-2016-15 \\
\end{flushright}

\vspace{5mm}

\begin{center}

{\Large\textbf{Flat  structures on the  deformations\\ of
  Gepner chiral rings
}}

\vspace{5mm}

\vspace{5mm}

{\large Alexander Belavin$^{\;a,c}$ and Vladimir Belavin$^{\;b,c,d}$}

\vspace{0.5cm}

\textit{$^a$ L.D. Landau Institute for Theoretical Physics, \\
 Akademika Semenova av., 1-A, \\ Chernogolovka, 142432  Moscow region, Russia}

\vspace{0.5cm}

\textit{$^{b}$I.E. Tamm Department of Theoretical Physics, \\P.N. Lebedev Physical
Institute,\\ Leninsky av., 53, 119991 Moscow, Russia}

\vspace{0.5cm}

\textit{$^{c}$Department of Quantum Physics, \\ 
Institute for Information Transmission Problems, \\
 Bolshoy Karetny per. 19, 127994 Moscow, Russia}

 \vspace{0.5cm}

\textit{$^{d}$Moscow Institute of Physics and Technology, \\
Dolgoprudnyi, 141700 Moscow region, Russia}

\vspace{0.5cm}

\thispagestyle{empty}


\end{center}
\begin{abstract}

We propose a simple method for  the computation of the  flat coordinates and Saito primitive forms on Frobenius manifolds of the deformations of Jacobi rings associated with isolated singularities.
The method is based on using a conjecture  about integral representations for the flat coordinates  and on the Saito cohomology theory. This reduces the computation to a simple linear problem. 
We consider the case of the deformed Gepner chiral rings.  
The knowledge of the flat structures  of Frobenius manifolds  can be used for exact solution
 of the models of the topological conformal field theories  corresponding to these chiral rings.

\end{abstract}
\newpage

\vspace{10mm}

\section{Introduction}
In this paper we study flat structures \cite{Saito} on Frobenius manifolds (FMs) \cite{Dub} arising on the deformations of isolated singularities. The physical interest to the singularity theory  \cite{Arnold} is motivated by its connection \cite{LVW,Mart}  with the chiral rings of $\mathcal{N}=2$ superconformal field theories (SCFTs).

The models of $\mathcal{N}=2$ SCFT, classified and explicitly constructed in \cite{KS}, are important due to their role
in string theory. It was shown in \cite{Gep} that $\mathcal{N}=2$ superconformal symmetry on the world sheet after compactifying six of the ten dimensions  is the necessary condition for the space-time supersymmetry.
The requirement of $\mathcal{N}=2$ supersymmetry in the compact sector is equivalent \cite{Gep2} to the geometrical condition\cite{CHSW} of the compactification on Calabi-Yau (CY) manifolds.
The massless sector of string theory corresponds to the set of chiral fields of  $\mathcal{N}=2$ SCFT
and after Witten's  twist\cite{W} it is described by the topological CFT (TCFT).
The properties of the Lagrangian of the sigma model describing this massless sector are completely  defined in terms of the CY manifold and its  mirror pair \cite {COGP,BCOFHJQ}. 
 
The FM  structure in TCFT was initially discovered in \cite{DVV} and later mathematically developed  in \cite{Dub}.  In \cite{DVV} it was shown that the exact solution of TCFT, that is  the computation of the correlation functions,  is reduced to the computation of the flat coordinates on  the corresponding FM. This is a specific property of 
$\hat{U}(2)_k \times \hat{U}(1)_1/ \hat{U}(1)_{k+1}$ models considered in \cite{DVV}. In the general case of $\mathcal{N}=2$ SCFT, associated with some isolated singularity one needs also to know the Saito  primitive form \cite{Saito}.   
      
It was established recently \cite{BSZ,BT,Tarn,BDM,VB,BB1,SPO,BYR,Belavin:2014xya} that the knowledge of FM structure allows to find exact solution
of a special class of models of non-critical string theory (or models of Minimal Liouville gravity) 
\cite{P,KPZ,AlZ,BAlZ}. It is natural to assume that the same approach can also help 
to solve the $\mathcal{W}$ generalizations of Liouville gravity.
The solution of the Douglas  string equation  \cite{Douglas}, required for the construction of the free energy, takes simple form only in the flat coordinates on the associated FM \cite{VB}. Otherwise the integral representation of the free energy is too complicated to be of any practical use.

The explicit form of the flat coordinates is important also for the solution of the TCFT models related to the Gepner chiral rings \cite{Gep1}.
It was shown   in \cite{Gep1} that these models, described by the   Kazama--Suzuki (KS) cosets
 $A(N,k)=\hat{U}(N)_k \times \hat{U}(N-1)_1/ \hat{U}(N-1)_{k+1}$,  are connected with a certain class of 
 isolated singularities.  The  super-potentials in this case is
\be 
W_0(q_1,...,q_{N-1})=\frac{1}{N+k}\sum_{i=1}^{N-1} q_i^{N+k}\;. 
\ee   

Thus, the computation of the flat coordinates and the Saito primitive form 
for the Frobenius manifolds represents  a significant part of the solution of the above mentioned problems. 

In this paper we suggest a new simple method for computing flat coordinates on the FM arising on the deformations of the isolated singularities. This problem was discussed earlier in \cite{Saito, Noumi, Blok, Losev,  BGK}. 
Our approach is based on  the conjecture  about integral representations for the flat coordinates  and on the Saito cohomology theory \cite{Saito}. This method allows to reduce the problem of the computation of the flat coordinates and the Saito primitive form \cite{Saito} to a linear problem. 
Namely, we show that the coefficients of the decomposition of the flat coordinates in a sum of monomials, constructed from the deformation parameters, are simply connected with the coefficients of the decomposition 
of the  dual to these monomials differential forms in a sum over basis elements  
in the space of the Saito cohomologies.  

We apply this approach  to the computation of  the flat coordinates  and the primitive form for Gepner chiral rings in the particular case of $A(3,k)=\hat{U}(3)_k\times \hat{U}(2)_1/\hat{U}(2)_{k+1}$ models.
This series is interesting because, unlike previously studied $A(2,k)$ series, in this case the marginal and irrelevant deformations are present in the spectrum. We demonstrate the general construction
in the particular case $k=3$ and find explicit formulae  for the flat coordinates and the primitive form
in terms of hypergeometric functions. 
The linear system for the computation of the flat coordinates together with the conjecture about the integral representation for them is the main result of the present paper.

The rest of this paper is organized as follows. In section \bref{sec1:FM} we discuss the Jacobi rings and recall some known facts  about Frobenius manifolds and their relations with the Saito theory.  
In section \bref{sec:genconstr}  we  develop our method for computing the flat coordinates and discuss the problem of finding the primitive form.  We formulate here the conjecture  about the integral representation for the flat coordinates for 
isolated singularity, whose Jacoby ring includes the marginal deformations in addition to the relevant ones.
 Further, we  describe the linear system for the computation of the flat coordinates in terms of the deformation parameters. 
 In section \bref{GepnerChirRing} we implement the proposed procedure to calculate the flat structure
of the FM associated with the Gepner chiral rings and  present the  details of the computation for the
 $SU(3)_3$  case. In section \bref{sec:conclusion} we conclude by summarizing and discussing some future perspectives.

\section {Singularities, Jacobi rings and Frobenius manifolds} 
\label{sec1:FM}

Here we recall the main facts about  the connection between Frobenius manifolds and topological conformal field theories relevant for our further discussion.

Dijkgraaf, Verlinde, and Verlinde (DVV) have found \cite {DVV} exact solution for the  minimal models of TCFT.  These models  are Landau-Ginzburg models whose superpotential  
\be
W_0(x)=\frac{1}{k+2}x^{k+2}\;.
\ee
The correlation functions in these models are defined by the free energy 
(also called frobenius prepotential)  $\mathcal{F}(s^1,...s^{k+1})$, 
where $s^{\mu}$ are the coupling constants of the theory and  $s^{1}$ corresponds to the unity operator.
The perturbed three-point function 
\be
C_{\alpha\beta\gamma}(s)=\langle\Phi_\alpha\Phi_\beta\Phi_\gamma
\exp{{\sum_{\sigma} s^\sigma\int d^2 z\Phi_\sigma^{1,1}(z)}}\rangle\;,
\ee
is given by
\be\label{strconst}
C_{\alpha\beta\gamma}=\frac{\partial^3 \mathcal{F}}{\partial s^\alpha\partial s^\beta\partial s^\gamma}\;.
\ee
It has been shown in \cite{DVV} that because of remarkable  properties of TCFT  all the information about arbitrary $n$-point correlation functions is encoded in the two basic objects: 
the three- and the two-point functions. The perturbed two-point function $\eta_{\alpha\beta}$ given explicitly by

\be
\eta_{\alpha\beta}=C_{1\alpha\beta}=
\langle\Phi_\alpha\Phi_\beta \exp{{\sum_{\sigma} s^\sigma\int d^2 z\Phi_\sigma^{1,1}(z)}}\rangle\;,
\ee
where $\Phi_\alpha(z)$ is a chiral field and $\Phi_\sigma^{1,1}(z)$ is its superpartner. 

Using  CFT Ward identities it has been proven in \cite{DVV} that $\eta_{\alpha\beta}$ and $C_{\alpha\beta\gamma}(s)$
should satisfy to the following set of the constraints
\be\label{FM1}
\p_\delta C_{\alpha\beta\gamma}=\p_\alpha C_{\delta\beta\gamma}\;,
\ee
\be\label{assoc}
C_{\alpha\beta}^\rho C_{\rho\gamma}^\delta=
C_{\alpha\rho}^\delta C_{\beta\gamma}^\rho\;.
\ee
\be\label{flat}
\partial_\alpha\eta_{\mu\nu}=0\;,
\ee
\be\label{FM4}
C_{\mu\nu\lambda}=C_{\nu\mu\lambda}=C_{\nu\lambda\mu}\;.
\ee
Here, in the second relation (associativity condition), raising indices is performed by  the inverse tensor $\eta^{\lambda\rho}$. From the eqs.(\ref{FM1}--\ref{FM4}) it follows that $C_{\mu\nu}^{\lambda}$ is a structure constant of some Frobenius algebra\footnote{We recall that Frobenius algebra is  a commutative and associative algebra with unity such that the multiplication low is agreed in a natural way with the definition of the  paring  \eqref{FM4}. Frobenius manifold represents a family of Frobenius algebras depending on a number of parameters. For more details see, e.g., \cite{Dub}.}
In particular, one of the main properties of the Frobenius algebra, eq.\eqref{strconst}, follows from \eqref{FM1} and\eqref{FM4}.
These relations mean that we are dealing with a Riemannian manifold, provided with some additional algebraic structure, which is compatible with the Riemannian one.  The parameters $s^\alpha$ play role  of the coordinates on this manifold and $\eta_{\rho\nu}$ defines its metric. 
We note that the third relation \eqref{flat} states that this  manifold is flat.   
In fact, the relations (\ref{FM1}--\ref{FM4}) define the structure of the Frobenius underlying TCFT.

\subsection{Dubrovin axioms}

 The parameters $s_\mu$ define
a special choice of {\it flat coordinates} on $M$. The metric $\eta^{\mu\nu}$ in these coordinates is constant.
In some general coordinate frame  $t=\{ t^\mu\}$  the Frobenius manifold axioms look like
\be\label{FM1g}
\nabla_\delta \widetilde C_{\alpha\beta\gamma}=\nabla_\alpha \widetilde C_{\delta\beta\gamma}\;,
\ee
\be\label{FM3g}
\widetilde C_{\alpha\beta}^\rho \widetilde C_{\rho\gamma}^\delta=
\widetilde C_{\alpha\rho}^\delta \widetilde C_{\beta\gamma}^\rho\;,
\ee
\be\label{FM4g}
\widetilde C_{\mu\nu\lambda}=\widetilde C_{\nu\mu\lambda} =\widetilde C_{\nu\lambda\mu}\;,
\ee
\be
R_{\mu\nu\lambda\sigma}[g_{\alpha\beta}]=0\;,
\ee
where $\nabla_\mu$ is the covariant derivative and $R_{\mu\nu\lambda\sigma}$ is Riemann  curvature tensor 
and  the structure constants in the new coordinates are denoted 
by $\widetilde C_{\alpha \beta}^\gamma({t})$.

It has been proven  in \cite{DVV} for the case of the  minimal models that  the constants $ C_{\mu\nu}^\lambda(s)$ are the structure constants in the flat coordinates of the Fobenius manifold connected with the deformed isolated singularity 
\be
W=W_0+\sum_{\mu=1}^{k+1}t^{\mu}e_\mu(x)\;, 
\ee
where $e_\mu(x)=x^{\mu-1}$ and the deformation parameters $t^{\mu}$ represent some coordinates
 on the FM. The parameter $t^{1}$ here and in what follows corresponds to the unity operator $e_1$ of the Jacobi ring.

In the $t$-frame the structure constants are defined  by 
\be\label{ee}
e_\mu e_\nu=\widetilde C_{\mu\nu}^\lambda (t) e_\lambda \bmod {\partial W}/{\partial x} \;.
\ee
The metric is
\be\label{metric_t}
g_{\mu\nu}=\underset{x=\infty}{\text{Res}} \,\frac{e_\mu e_\nu \Omega}{{\partial W}/{\partial x}}
\;,
\ee
where $\Omega$ is the Saito primitive form.
 In the case of  minimal models 
\be
\Omega= d x\;.
\ee

According to the statement formulated above, in order to solve the model we need to know the flat coordinates $s^\mu(t)$ on the FM as functions of the deformation parameters.
If the metric is flat, the flat coordinates $s^\mu(t)$ are given by the solutions of the equation
\be
\nabla_\alpha \nabla_\beta s^\mu =\Gamma_{\alpha\beta}^\gamma \nabla_\gamma s^\mu\;,
\ee
where $\Gamma_{\alpha\beta}^\gamma$ is Levi-Civita connection for $g_{\alpha\beta}(t)$. 

The free energy $\mathcal{F}$ is defined from 
\be\label{C-F}
C_{\mu\nu\lambda}=\frac{\partial^3 \mathcal{F}}{\partial s^\mu \partial s^\nu \partial s^\lambda}\;,
\ee
and
\be\label{C-res}
C_{\mu\nu\lambda}\frac{\partial s^\mu}{\partial t^\alpha}\frac{\partial s^\nu}{\partial t^\beta}\frac{\partial s^\lambda}{\partial t^\gamma}=\underset{x=\infty}{\text{Res}} \,\frac{e_\alpha e_\beta e_\gamma \Omega}{{\partial W}/{\partial x_1}\cdots{\partial W}/{\partial x_n}}
\;.
\ee
In fact, the eqs. \eqref{C-F} and \eqref{C-res} give us the exact solution of the minimal models of TCFT as soon as we know
the explicit expression for the flat coordinates as functions of the deformation parameters.

\subsection{K.~Saito primitive form}
Let $W_0(x_1,...,x_n)$ be a polynomial
associated with an isolated quasihomogenous singularity
with integer weights $[x_i]=\rho_i$ and $[W_0]=d$
\be\label{homogen}
W_0(\Lambda^{\rho_i} x_i)=\Lambda^d W_0(x_i)\;. 
\ee
By definition, the corresponding Jacobi ring is given by
\be
R_0=\mathbb{C}[x_1,...,x_n] /\left\{\frac{\partial W_0[x]}{\partial x_i}\right\}\;.
\ee
 In the analogous way as in the case of one variable,  the {\it versal deformation} of the ring $R_0$ is given by
\be
W(x_i,t_\alpha)=W_0(x_i)+\sum_{\alpha=1}^M t^\alpha e_\alpha(x)\;, 
\ee
 where $e_\alpha$, $\alpha=1,...,M$, represent a basis of
the corresponding deformed Jacobi ring and $M$ is the Milnor number (the dimension of the Jacobi ring)
\be
R=\mathbb{C}[x_1,...,x_n]/ \left\{\frac{\partial W[x,t]}{\partial x_i}\right\}\;.
\ee
We denote the weights of the homogeneous basis elements 
$[e_\alpha]=\sigma_\alpha$. In our convention $e_1$ is the unity in the ring. The following dimensional classification will be relevant.  The elements $e_\alpha$ with $\sigma_\alpha<d$, $\sigma_\alpha=d$ and
$\sigma_\alpha>d$ are called respectively relevant, marginal and irrelevant deformations.

The generalization of the eq.\eqref{ee} is straightforward, 
while the analogue of \eqref{metric_t} is  
\be\label{metric_g_gen}
g_{\mu\nu}=\underset{x=\infty}{\text{Res}} \,\frac{e_\mu e_\nu \Omega}{\prod_i \partial_i W}
\;,
\ee
where the primitive form 
\be\label{omega}
\Omega(t, x)= \lambda(t^\alpha, x_i) d x_1\wedge...\wedge d x_n\;. 
\ee 
We will use the following important statement proved by M.~Saito \cite{M.Saito}.

{\bf Theorem}. {\it If $W_0$ is an isolated singularity, then there exists such a $n$-form  
$\Omega[x]$ or, equivalently, a function $\lambda(x,t)$, $[\lambda]=1$,   that the above defined $C_{\alpha\beta}^\gamma$ and $g_{\alpha\beta}$  satisfy all FM axioms (\ref{FM1}--\ref{FM4}).}

We note that in the DVV case, $\lambda(x,t)=1$. In the general case, if there are marginal and/or irrelevant deformations, $\lambda(x,t)\neq 1$.

From this theorem it follows that there exist flat coordinates  $s^\mu(t)=t^\mu+\mathcal{O}(t^2)$, such that  $[s^\mu(t)]=[t^\mu]$. 
Additionally, the coordinate $s^1(t)$ is fixed by the condition that it represents a coupling of the unity operator, that is 
\be 
e_1= \frac{\partial W}{\partial s^1}\;.
\ee
As a consequence of this definition, one gets
\be \label{e1}
\frac{\partial s^{\mu}}{\partial t^1}=\delta_{\mu,1}\;,
\ee
which follows from the simple argument
\be 
e_1= \frac{\partial W}{\partial t^1}=
\frac{\partial W}{\partial s^{\mu}}\frac{\partial s^{\mu}}{\partial t^1}=
\frac{\partial s^{\mu}}{\partial t^1}e_{\mu}\;.
\ee
Eq.\eqref{e1}  imposes the strong constraints on the primitive form discussed below. In fact, they allow to compute explicitly the  primitive form simultaneously with the flat coordinates. 

\section{Approach for the computation of the flat coordinates}
\label{sec:genconstr}

\subsection{Integral representation for flat coordinates on FM}

  It was argued in \cite{BGK}, using the connection between the  Jacobi rings  of the deformed  singularities and the Gauss-Manin (GM) systems,
  that the flat coordinates  $s^{\mu}$ considered  as functions of the parameters $t^{\mu}$ accept a simple integral representation.  It was verified in \cite{BGK} for the case of simple singularities, that is when  the  Jacobi ring includes only relevant deformations.

Here we conjecture that similar integral representations exist also in general case when the Jacobi ring includes the marginal deformations and probably irrelevant ones.
The integration contours should be closed and  chosen  to belong to  a basis in the homologies $H_n$ of $\mathbb{C}^n \setminus\{W=0\}$
\be\label{sm_int}
s^{\mu}(t^{\alpha})= \sum_{m_{\alpha}\in \Sigma_{\mu}} 
\left(\int_{\gamma_{\mu}} \exp(W_0(x)) \prod_{\alpha} e_{\alpha}^{m_{\alpha}} \Omega\right) 
\prod_{\alpha} \frac{t_{\alpha}^{m_a}}{m_{\alpha}!}\;,
\ee
where non-negative integers $m_\alpha\in  \Sigma_\mu$ if
\footnote{For convenience reason in \eqref{sm_int} and in what follows in order to avoid additional brackets we replace sometimes superscripts by subscripts $t_\alpha\equiv t^\alpha$ and  $s_\alpha\equiv s^\alpha$. This operation has nothing to do with raising and lowering indices by the metric.}
  
\be\label{balance}
\sum_\alpha m_\alpha [t^\alpha]=[s^\mu]\;.
\ee
In what follows we focus on the case where all elements $e_\alpha$ of the basis of the Jacobi ring are relevant or marginal. In this case $\Omega(t,x)$ is given by \eqref{omega}, $\lambda(t)$ is a function of only those   parameters $t_\alpha$ which correspond to the marginal operators and does not depend on $x_i$.
Therefore \eqref{sm_int} can be rewritten in the following form 
\be\label{sm_int_new}
s^{\mu}(t^{\alpha})= \lambda(t)\sum_{m_{\alpha}\in \Sigma_{\mu}} 
\left(\int_{\gamma_{\mu}} \exp(W_0(x)) \prod_{\alpha} e_{\alpha}^{m_{\alpha}} dx \right) 
\prod_{\alpha} \frac{t_{\alpha}^{m_\alpha}}{m_{\alpha}!}\;.
\ee
On the other hand, from the dimensional arguments it is obvious that in general form
\be
s^\mu(t^1,...,t^M)=t^\mu +\sum_{m_\alpha\in \Sigma_\mu} C_\mu(\vec{m}) \prod_{\alpha=1}^M 
\frac{t_\alpha^{m_\alpha}}{m_\alpha !}\;.
\ee
In what follows we show how to compute the coefficients $ C_\mu(\vec{m})  $.

\subsection{Saito cohomologies and exact expressions for Flat coordinates }
For explicit computations of the integral in \eqref{sm_int} we will use the following two facts.

Firstly, two following  integrals are equal
\be
      \int_\gamma \exp(W_0(x)) P_1(x) dx=\int_\gamma \exp(W_0(x)) P_2(x) dx
      \;,
      \ee
      if
      \be 
       P_1(x) dx=P_2(x) dx + D_{W_0} U\;,
       \ee
where the K.~Saito differential\footnote{Here $d$ is the ordinary de Rham differential. We note that $D_{W_0}^2=0$.}
\be 
D_{W_0}=d+dW_0\;,
\ee
and $U$ is some  $(n-1)$-form
\be
U=\sum_{i=1}^n (-1)^{i-1} A_i dx_1\wedge ...\wedge dx_{i-1}\wedge dx_{i+1} \wedge... \wedge dx_n
\ee
and contour $\gamma$ is closed.

Secondly, the Saito differential defines  cohomologies on the space of the differential forms. The dimension $H^n$-- Saito  cohomology group 
of the highest degree  is equal $M$, i.e., the dimension of the Jacoby ring of the isolated singularity $W_0$.
As a basis $H^n$  can be choosen the $n$-forms $e_\mu dx_1 \wedge...\wedge dx_n$  or $e_\mu dx$ in short, where $e_\mu$ is a basis  of the Jacobi ring. Since the forms $P(x) dx_1 \wedge...\wedge dx_n$ belong to $H^n$ they can be decomposed in this basis
\be 
P(x) dx= \sum_{\mu} C_\mu e_\mu dx+D_{W_0} U\;.
\ee
Such an expansion exists and is unique.
 We will use this fact  to find the coefficients $ C_\mu(\vec m)$ from the equation
\be\label{proda}
\prod_{\alpha} e_\alpha^{m_\alpha} dx=\sum_\mu C_\mu(\vec m) e_\mu dx+ D_{W_0} U\;.
\ee
This equation can be written more explicitly as
\be
\prod_{\alpha} e_\alpha^{m_\alpha}=\sum_\mu   C_\mu(\vec m) e_\mu+ \sum_{i=1}^n A_i(x) \frac{\partial W_0}{\partial x^i}
+ \sum_{i=1}^n  \frac{\partial A_i(x)}{\partial x^i}\;.
\ee
Assuming that the basis contours $\gamma_\mu$ are chosen to be dual with  elements $e_\mu dx$ in the basis of  the cohomology space, i.e.,
           \be  \int_{\gamma_\mu} e_\nu dx = \delta_{\mu,\nu}\;,
\ee
and substituting the RHS in \eqref{proda} into the integrals in \eqref{sm_int} we arrive to explicit expression for the flat coordinates 
\be  
 s^\mu(t^\alpha)= \sum_{m_\alpha} C_\mu(m_\alpha) \prod_{\alpha} \frac{t_\alpha^{m_\alpha}}{(m_\alpha)!}\;,
\ee
where the coefficients $C_\mu(\vec{m})$  are unique solution of the system of linear equations \eqref{proda}.

If the isolated singularity is homogeneous, the $(n-1)$-form $U$ can be represented as a sum of 
homogeneous terms $\sum_{l=0}^{L} U_l$ and the equation \eqref{proda} takes the form
\begin{align}\label{rec1}
&d W_0\, U_0 =\prod_\lambda e_\lambda^{m_\lambda} \,dx\;,\\\label{rec2}
&d W_0\, U_l=- d U_{l-1}\;,\\\label{rec3}
&\sum_\mu C_\mu(\vec{m}|n) e_\mu \, d x=-d U_L\;,
\end{align}
where $l=1,...,L$ and 
$L=\sum m_\alpha -1$.

In the next section we will demonstrate how this procedure works in the case of the Gepner rings.

\section{Computation of the flat coordinates for Gepner chiral rings}
\label{GepnerChirRing}
\subsection{Gepner chiral rings and their deformations}

Chiral rings considered by Gepner are related to the KS cosets 
$\hat{U}(N)_k \times \hat{U}(N-1)_1/ \hat{U}(N-1)_{k+1}$.
Important property of these rings is that their deformations possess a structure  of Frobenius manifolds.
It  follows  \cite{Gep1} from the fact that these chiral rings are connected with the isolated homogeneous  singularity
\be
W_0(x)=\frac{1}{N+k}\sum_{i=1}^{N-1} q_i^{N+k}\;,
\ee
and are isomorphic to the Jacoby rings
\be\label{R0}
R_0(N,k)=\mathbb{C}[x_1,...,x_{N-1}]\big/\{\partial_1 W_0[x]...\partial_{N-1} W_0[x]\}\;,
\ee
where $x_1,...,x_{N-1}$ are elementary symmetric polynomials of $q_1,...,q_{N-1}$:
\begin{align}
x_1=\sum_{i=1}^{N-1} q_i\;,\quad
x_2=\sum_{i<j}^{N-1} q_i q_j\;,\quad
...\;,\quad
x_{N-1}=q_1...q_{N-1}\;.
\end{align}
Note that the dimension of the ring is
\be\label{dim}
\dim R_0(N,k)=\frac{(N+k-1)!}{k!(N-1)!}\;.
\ee
Taking into account symmetric property of the ring space it is clear that Schur polynomials 
\be\label{schur}
S_{\lambda}[q_1,...,q_{N-1}]=\frac{\det q_i^{N+\lambda_j-j}}{\det q_i^{N-j}}
\ee
can be chosen as
 a natural homogeneous basis in $R^0(N,k)$. 
We remind that  $\lambda$ in \eqref{schur} stands for the Young diagram $\lambda=(\lambda_1,....,\lambda_{N-1})$, with $\lambda_1\geq \lambda_2\geq...\geq\lambda_{N-1}\geq0$ and $0\leq\lambda_1\leq k$.

\subsection{Computation of the flat coordinates for $SU(3)_3$ chiral ring }
For $k\geq 3$ the consideration of the models associated with the Gepner chiral rings \eqref{R0} become more involved because marginal deformations appear in the spectrum.
Let us consider the particular case $k=3$ where
\be\label{W0k3}
W_0(x_1,x_2)=\frac{1}{6}( q_1^6+q_2^6 )\;. 
\ee
According to \eqref{dim} the dimension of the ring space $\dim R_0(3,3)=10$. The basis elements are enumerated by Young diagrams restricted to a rectangle of size
$k\times (N-1)$ with $k=3$, $N=3$. For convenience we will use the following identification
\begin{align}
e_1\equiv e_{\varnothing}=1\;, \quad 
e_2\equiv e_{\tableau{1}}=q_1+q_2\;, \quad
e_3\equiv e_{\tableau{1 1}}=q_1 q_2\;. \quad
e_4\equiv e_{\tableau{2}}=q_1^2+q_1 q_2+q_2^2\;, \quad \nonumber\\
e_5\equiv e_{\tableau{2 1}}=q_1 q_2(q_1+q_2)\;, \quad
e_6\equiv e_{\tableau{3}}=q_1^3+q_1^2 q_2+q_1 q_2^2+q_2^3\;,\quad
e_7\equiv e_{\tableau{2 2}}=q_1^2q_2^2\;,\quad\nonumber\\
e_8\equiv e_{\tableau{3 1}}=q_1 q_2(q_1^2+q_1 q_2+q_2^2)\;,\quad
e_9\equiv e_{\tableau{3 2}}=q_1^2 q_2^2(q_1+q_2)\;,\quad
e_{10}\equiv e_{\tableau{3 3}}=q_1^3 q_2^3\;.\quad
\end{align}

\noindent After the versal deformation the potential  reads
\be\label{R33}
W(q_1,q_2,t)=\frac{q_1^6+q_2^6}{6} +\sum_{l=1}^{10} t^{l} e_l   \;.
\ee
Note that in this particular case there is only one marginal deformation which is presented by $e_{10}$, all other perturbations $e_{1},...,e_{9}$ in \eqref{R33} are relevant. Therefore we have to allow 
the primitive form \eqref{omega} to be a function of the dimensionless coupling $t_{10}$.
The coordinates in the constant metric coordinate system are given by 
the following general expression
\be
s^\mu(t^\alpha)=\lambda(t_{10})\sum_{\{m_\lambda\}\in \Sigma_\mu} \sum_{n=0}^\infty C_\mu(m_\lambda | n) \prod_{1\leq\lambda< 10} 
\frac{t_\lambda^{m_\lambda}}{m_\lambda !} \frac{t_{10}^n}{n!}\;,
\ee
where $\Sigma_\mu$ is defined in \eqref{balance}. In accordance with our main conjecture \eqref{sm_int} the coefficients $ C_\mu(m_\lambda | n)$ are fixed by the following condition 
\be\label{cond}
\prod_{\lambda} e_\lambda^{m_\lambda} e_{10}^n \,\Omega_0=\sum_\mu C_\mu(m_\lambda|n) e_\mu\, \Omega_0 +D_{W_0} U\;,
\ee
where 
\be
\Omega_0=(q_1-q_2)dq_1 \wedge d q_2\;,
\ee
and the sum in \eqref{cond} is over those $\mu$, which are compatible  with the dimensional restriction \eqref{balance} for given set of $m_\lambda$. The 1-form $U$ can be conveniently represented as 
\be
U=A(q_1,q_2) d q_1+B(q_1,q_2) d q_2\;,
\ee
such that $B(q_1,q_2)=A(q_2,q_1)$. 
More explicitly we find
\be
\prod_{\lambda} e_\lambda^{m_\lambda} e_{10}^n (q_1-q_2)=\sum_\mu C_\mu(m_\lambda|n) 
e_\mu (q_1-q_2) +\frac{\partial B}{\partial q_1}-\frac{\partial A}{\partial q_2}+A q_2^5- B q_1^5\;.
\ee
Taking into account  the argumentation of section \bref{sec:genconstr}, we can transform this relation into the set of recurrence relations for its decomposition into homogeneous pieces  
\be 
A=\sum_{s=0}^L A_s\,, \quad B=\sum_{s=0}^L B_s\;,
\ee 
where $L$ is defined from the condition
\be
[\prod_\lambda e_\lambda^{m_\lambda} e_{10}^n]-[e_\mu]=6 L\;.
\ee
The first relation of the recurrence prescription (\ref{rec1}-\ref{rec3}) reads 
\be
 \prod_{\lambda} e_\lambda^{m_\lambda}(q_1-q_2)= A_0 q_2^5-B_0 q_1^5\;.
\ee
The second one gives
\be 
\frac{\partial  A_{s-1}}{\partial q_2} - \frac{\partial  B_{s-1}}{\partial q_1}=
A_s q_2^5 -B_s q_1^5,\qquad 1\leq s\leq L\;.
\ee
Finally, from the last one we find
\be 
\sum_\mu C_\mu(m_\lambda |n) e_\mu(q_1,q_2)=\frac{\partial  A_{L}}{\partial q_2} - \frac{\partial  B_{L}}{\partial q_1}\;.
\ee
These iterative relations allow to find all the coefficient $A_s, B_s$ and to express finally the desired 
$C_\mu(m_\lambda |n)$ in terms of the highest coefficients $A_L, B_L$.

\subsection{Explicit expressions for the flat coordinates for  $t_{10}=0$}

Now we want to find the flat coordinates. Our first step is to consider the situation with the marginal perturbation
switched off. In this case the most general coordinate transformation which is compatible with the requirement of quasihomogeneity condition \eqref{homogen} can involve only the following contributions
\begin{align}
s_{10}:{}&\{t_{10}\}\;,\\ 
s_9:{}&\{t_9\}\;, \\
s_8:{}&\{t_8, t_9^2\}\;, \\
s_7:{}&\{t_7, t_9^2\}\;,\\ 
s_6:{}&\{t_6, t_8 t_9, t_7 t_9, t_9^3\}\;, \\
s_5:{}&\{t_5, t_8 t_9,  t_7 t_9, t_9^3\}\;, \\
s_4:{}&\{t_4, t_6 t_9, t_5 t_9,  t_8 t_9^2, t_7 t_9^2, t_8^2, t_7^2, t_7 t_8, t_9^4\}\;,\\
s_3:{}&\{t_3,  t_6 t_9,  t_5 t_9, t_8 t_9^2, t_7 t_9^2, t_8^2, t_7^2, t_7 t_8, t_9^4\}\;, \\
s_2:{}&\{t_2, t_4 t_9, t_7^2 t_9, t_7 t_8 t_9, t_8 t_9^3, t_7 t_9^3, t_9^5, t_3 t_9,
t_6 t_8, t_6 t_7, t_5 t_8, t_5 t_7, t_6 t_9^2, t_5 t_9^2, t_8^2 t_9\}\;,\\
s_1:{}&\{t_1, t_2 t_9, t_5 t_6, t_6 t_8 t_9, t_6 t_7 t_9,  t_5 t_8 t_9, t_5 t_7 t_9, t_6 t_9^3, t_5 t_9^3, t_8^3, t_7^3, t_7 t_8^2, t_4 t_8, t_7^2 t_8, t_8^2 t_9^2, \nonumber\\
& t_7^2 t_9^2, t_7 t_8 t_9^2, t_8 t_9^4, t_7 t_9^4, t_9^6, t_4 t_7, t_3 t_8, t_3 t_7, t_4 t_9^2, t_3 t_9^2, t_6^2, t_5^2\}\;.
\end{align}
Here each monomial listed in the right hand side comes in the expressions of the corresponding flat coordinate with a coefficient
depending on $t_{10}$.
For $t_{10}=0$, our approach described in the previous section gives the following explicit results 
\begin{align}
s_{10}={}&t_{10}\;,\\ 
s_9={}&t_9\;, \\
s_8={}&t_8+ t_9^2\;, \\
s_7={}&t_7\;,\\ 
s_6={}&t_6+t_8 t_9+ t_7 t_9+\frac{5}{6} t_9^3\;, \\
s_5={}&t_5+2 t_8 t_9+\frac{8}{3} t_9^3\;, \\
s_4={}&t_4+ t_6 t_9+ t_5 t_9+\frac{1}{2} t_8^2+ t_7 t_8+2 t_7 t_9^2+4 t_8 t_9^2+\frac{7}{2} t_9^4\;,\\
s_3={}&t_3+2 t_6 t_9+3  t_8 t_9^2+3 t_7 t_9^2+t_8^2+ 2 t_9^4\;, \\
s_2={}&t_2+\frac{63 t_9^5}{8}+\frac{7}{2} t_7 t_9^3+\frac{21}{2} t_8 t_9^3+3 t_5 t_9^2+\frac{3}{2} t_6 t_9^2+\frac{3}{2} t_7^2 t_9+\nonumber\\
&+3 t_8^2 t_9+t_4 t_9+3 t_7 t_8 t_9+t_6 t_7+t_5 t_8+t_6 t_8\;,\\
s_1={}& t_1+\frac{119 t_9^6}{15}+\frac{28}{3} t_7 t_9^4+14 t_8 t_9^4+\frac{7}{3} t_5 t_9^3+\frac{13}{3} t_6 t_9^3+\frac{13}{2} t_8^2 t_9^2+t_3 t_9^2+t_4 t_9^2+\nonumber\\& 
+2 t_5 t_7 t_9+7 t_7 t_8 t_9^2+
2 t_5 t_8 t_9+2 t_6 t_8 t_9+\frac{t_7^3}{3}+\frac{t_8^3}{3}+t_7 t_8^2+t_5 t_6+t_4 t_8-\frac{t_6^2}{2}\;.
\end{align}

\subsection{Explicit expressions for the flat coordinates for $t_{10}\neq0$}
When $t_{10}\neq 0$ the flat coordinates are series of $t_{10}$.
Let us demonstrate the general idea of the previous section on the following examples for $k=3$.

In our first example we compute the contribution to $s^1$ which contains $t^1$. It has the following form
\be 
{s}^1=\lambda(t_{10}) f_{1}(t_{10}) t^1 + ...\;, 
\ee
where dots stand for  possible contributions of other $t^{\nu}$ with $\nu \neq1$.
Our goal now is to find $f_{1}(t_{10})$.
\be 
f_{1}(t_{10})=\sum_n C(n)  \frac{t_{10}^{2n}}{(2n)!} \;.
\ee
As explained above we compute the coefficients $C(n)$ from the equation 
\be
e_1 e_{10}^{2n} \Omega = C(n) \Omega + D_{W_0} U\;.
\ee
Since $e_1=1$, we have to solve
\be
e_{10}^{2n} (q_1-q_2) = C(n) (q_1 -q_2) +
\frac{\partial B}{\partial q_1}-\frac{\partial A}{\partial q_2}+A q_2^5- B q_1^5\;.
\ee
It is convenient to derive the following recurrence relation for the coefficients
\be
C(n+1)=P(n)C(n)\;.
\ee
Since $e_{10}=q_1^3 q_2^3$, we get
\be
q_1^{6n +6} q_2^{6n+6}(q_1-q_2)=q_1^{6n} q_2^{6n}(q_1-q_2)+...\;,
\ee
which is to be presented in the form
\be
q_1^{6n +7} q_2^{6n+6}-q_1^{6n +6} q_2^{6n+7}=A q_2^5-Bq_1^5+
\frac{\partial B}{\partial q_1}-\frac{\partial A}{\partial q_2}+
P(n)[q_1^{6n +1} q_2^{6n}-q_1^{6n +6} q_2^{6n+1}]\;.
\ee
The coefficients are easily found $A_0=-q_1^{6n+6}q_2^{6n+2}$, $B_0=-q_1^{6n+2}q_2^{6n+6}$.
So that
\be
\frac{\partial A_0}{\partial q_2}-\frac{\partial B_0}{\partial q_1}=(6n+2)
[q_1^{6n +1} q_2^{6n+6}-q_1^{6n +6} q_2^{6n+1}]\;,
\ee
which in turn is to be equal to
\be
(6n+2)[q_1^{6n +1} q_2^{6n+6}-q_1^{6n +6} q_2^{6n+1}]=A_1q_2^5-B_1q_1^5\;,
\ee
so that  $A_1=(6n+2)q_1^{6n+1}q_2^{6n+1}$, $B_1=(6n+2)q_1^{6n+1}q_2^{6n+1}$.
The final condition is
\be
\frac{\partial A_1}{\partial q_2}-\frac{\partial B_1}{\partial q_1}=(6n+1)(6n+2)
[q_1^{6n +1} q_2^{6n}-q_1^{6n} q_2^{6n+1}]=P(n)[q_1^{6n +1} q_2^{6n}-q_1^{6n} q_2^{6n+1}]\;.
\ee
Hence,
\be
P(n)=(6n+1)(6n+2)\;,
\ee
and
\be
C(n)=\prod_{m=0}^{n-1}(6m+1)(6m+2)\;.
\ee
We conclude that
\be
f_{1}(t_{10})=1+\sum_{n=1}^{\infty}\prod_{m=0}^{n-1}(6m+1)(6m+2)\frac{t_{10}^{2n}}{(2n)!}=
{}_2F_{1}\ll\frac{1}{6},\frac{1}{3};\frac{1}{2}\bigg|9t_{10}^2\rr\;.
\ee
Taking into account the \eqref{e1} we obtain
\be
\lambda(t_{10})=\frac{1}{f_{1}(t_{10})}=\frac{1}{{}_2F_{1}\ll\frac{1}{6},\frac{1}{3};\frac{1}{2}\bigg|9t_{10}^2\rr}\;.
\ee

Our second example demonstrates  the computation of the flat coordinate 
\be
s_{10}=\lambda(t_{10}) t_{10} f_{10}(t_{10})=
\lambda(t_{10})\sum_{n=1}^\infty C(n) \frac{t_{10}^{2n+1}}{(2n+1)!}\;,
\ee
with $C(0)=1$. We have the following equations for the coefficients $C(n)$ 
\be
e_{10}^{2n+1}\Omega_0=C(n) e_{10}\Omega +D_{W_0} U\;,
\ee
and 
\be
C(n+1)=P(n)C(n)\;.
\ee
Or explicitly, 
\be
e_{10}^{2n+3}(q_1-q_2)=P(n)e_{10}^{2n+1}(q_1-q_2)+A q_2^5-B q_1^5+\frac{\partial B}{\partial q_1}
-\frac{\partial A}{\partial q_2}\;.
\ee
It follows that 
\begin{align}
U=U_0+U_1\;,\quad
A=A_0+A_1\;,\quad B=B_0+B_1\;,
\end{align}
where
\begin{align}
&A_0=-q_1^{6n+9}q_2^{6n+4}\;, \qquad\,\,\,\,\, B_0=-q_1^{6n+4} q_2^{6n+9}\;,\\
&A_1=(6n+4)q_1^{6n+3}q_2^{6n+5}\;,\quad B_1=(6n+4)q_1^{6n+5}q_2^{6n+3}\;,
\end{align}
and
\begin{align}
P(n)=(6n+4)(6n+5)\;. 
\end{align}
We find
\be
f_{10}(t_{10})=1+\sum_{n=1}^\infty \prod_{m=0}^{n-1}(6m+4)(6m+5) \frac{t_{10}^{2n+1}}{(2n+1)!}\;.
\ee
One can easily see that
\be
s_{10}(t_{10})= t_{10} \lambda(t_{10}) \, {}_2 F_{1} \ll\frac{2}{3},\frac{5}{6};\frac{3}{2}\bigg|9t_{10}^2\rr
= t_{10} \frac{{}_2F_{1}\ll\frac{2}{3},\frac{5}{6};\frac{3}{2}\big|9t_{10}^2\rr}
{{}_2F_{1}\ll\frac{1}{6},\frac{1}{3};\frac{1}{2}\big|9t_{10}^2\rr}\;\;.
\ee
Let us give another few examples of the explicit results for the flat coordinates when the final expressions are not too cumbersome 
\begin{align}
s_{9}(t)={}& t_{9} \frac{{}_2F_{1}\ll\frac{5}{6},\frac{1}{2};\frac{1}{2}\big|9t_{10}^2\rr}
{{}_2F_{1}\ll\frac{1}{6},\frac{1}{3};\frac{1}{2}\big|9t_{10}^2\rr}\;,\\
s_{8}(t)={}& t_{8} \frac{{}_2F_{1}\ll\frac{2}{3},\frac{1}{2};\frac{1}{2}\big|3t_{10}\rr}
{{}_2F_{1}\ll\frac{1}{6},\frac{1}{3};\frac{1}{2}\big|9t_{10}^2\rr}+
t_{9}^2\frac{ {}_2F_{1}\ll\frac{5}{3},\frac{1}{2};\frac{1}{2}\big|3t_{10}\rr}{{}_2F_{1}\ll\frac{1}{6},\frac{1}{3};\frac{1}{2}\big|9t_{10}^2\rr}\;,\\
s_{7}(t)={}& t_{7}\frac{ {}_2F_{1}\ll\frac{2}{3},\frac{1}{2};\frac{1}{2}\big|9t_{10}^2\rr}
{{}_2F_{1}\ll\frac{1}{6},\frac{1}{3};\frac{1}{2}\big|9t_{10}^2\rr}+
6 t_{9}^2 t_{10} \frac{ {}_2F_{1}\ll\frac{5}{3},\frac{1}{2};\frac{1}{2}\big|9t_{10}^2\rr}{{}_2F_{1}\ll\frac{1}{6},\frac{1}{3};\frac{1}{2}\big|9t_{10}^2\rr}\;.
\end{align}
The demonstrated above procedure, which allows to find explicit  expressions for the flat coordinates, the formula for the Saito primitive form $\Omega$
\be
\Omega=\lambda(t_{10})dx=\frac {dx}{{}_2F_{1}\ll\frac{1}{6},\frac{1}{3};\frac{1}{2}\bigg|9t_{10}^2\rr}\;,
\ee
together with the statement about the form of the flat coordinates, \eqref{sm_int}, \eqref{proda}, are the main results of this paper.

\section{Conclusion}
\label{sec:conclusion}

We have studied $\mathcal{N}=2$ CFT models, represented by cosets  $\hat{U}(N)_k \times \hat{U}(N-1)_1/ \hat{U}(N-1)_{k+1}$ for $N>2$. An important feature of these models is the presence of marginal and irrelevant perturbations.
The proposed method allows to systematically analyse such situations. 
It gives the recurrence formula for the flat coordinates and the method to construct the primitive form. This result gives the complete solution of the model under consideration.

Among the open questions we would like to mention the following.
Since the mathematical proof of the main conjecture is lacking, 
the direct computation of the flat coordinates, in the spirit of \cite{Klemm}, is desirable.
In the case where only relevant and marginal perturbations are presented, the check was performed in \cite{BB}. It is important to investigate the case of irrelevant perturbations.

Our results can be used in order to define the generating function of the physical amplitudes in the minimal models of 2d $\mathcal{W}$ gravity. Note that the generalization of the discrete  formulation of the matrix models possessing  $\mathcal{W}$ symmetry is not available  at the moment.
It is also worth noting that the analogue of the continuous approach, based on the explicit construction
of the correlators  of the chiral fields, is also absent.\footnote{See \cite{Belavin:2012uf} for some development in this direction.}

Finally, maybe the most interesting is the possibility 
to investigate a new class of Calabi-Yau geometries, using the approach discussed in this paper.

\vspace{7mm} 

\noindent \textbf{Acknowledgements.} We thank M.~Bershtein, D.~Gepner, B.~Feigin, Ya.~Kononov, L.~Spodyneiko and G.~Tarnopolsky for useful discussions. 
The work was performed with the financial support of the Russian Science Foundation (Grant No.14-12-01383).


%

\providecommand{\href}[2]{#2}\begingroup\raggedright

\endgroup

\end{document}